\documentclass[reprint,amsmath,aps,prb,footinbib]{revtex4-2}
\usepackage{graphicx,epstopdf,dcolumn,bm,mathtools,siunitx,multirow,hyperref,xcolor}
\newcommand{\mli}[1]{\mathit{#1}}
\begin{document}
\title{Spectral properties of size-invariant shape transformation}
\author{Alhun Aydin$^{1,2}$}
\email{alhunaydin@fas.harvard.edu}
\affiliation{$^{1}$Department of Physics, Harvard University, Cambridge, MA 02138, USA \\
$^{2}$Department of Physics, Ko\c{c} University, 34450 Sar\i yer, Istanbul, Turkey}
\date{\today}
\begin{abstract}
Size-invariant shape transformation is a technique of changing the shape of a domain while preserving its sizes under the Lebesgue measure. In quantum confined systems, this transformation leads to so-called quantum shape effects in the physical properties of confined particles associated with the Dirichlet spectrum of the confining medium. Here we show that the geometric couplings between levels generated by the size-invariant shape transformations cause nonuniform scaling in the eigenspectra. In particular, the nonuniform level scaling is characterized by two distinct spectral features: lowering of the first eigenvalue (ground state reduction) and changing of the spectral gaps (energy level splitting or degeneracy formation depending on the symmetries). We explain the ground state reduction by the increase in local breadth (i.e. parts of the domain becoming less confined) that is associated with the sphericity of these local portions of the domain. We accurately quantify the sphericity using two different measures: the radius of the inscribed $n$-sphere and the Hausdorff distance. Due to Rayleigh–Faber–Krahn inequality, the greater the sphericity, the lower the first eigenvalue. Then, level splitting or degeneracy, depending on the symmetries of the initial configuration, becomes a direct consequence of size-invariance dictating the eigenvalues to have the same asymptotic behavior due to Weyl law. Furthermore, we find that the ground state reduction causes a quantum thermal avalanche which is the underlying reason for the peculiar effect of spontaneous transitions to lower entropy states in systems exhibiting the quantum shape effect. Unusual spectral characteristics of size-preserving transformations can assist in designing confinement geometries that could lead to classically inconceivable quantum thermal machines. 
\end{abstract}
\maketitle
\section{Introduction}
Spectral geometry deals with the relationships between the spectrum of a differential operator and the geometry of the manifold on which it acts upon\cite{specbook1}. With the remarkable progress in nanoscience and nanotechnology in the last decades, understanding the spectra of finite-size systems has become crucial for the geometric design of nanostructures with enhanced physical properties\cite{baltes,pathria,mitin}. In quantum confined systems, the discrete energy spectra make the geometry dependence of the physical properties of materials prominent, which were negligible at macroscale\cite{bineker,Yoffe_2002,sismanmuller,Koksharov,McNamara_2010}. Quantum size effect is perhaps the best known of the geometric effects appearing at nanoscale\cite{rodun,qconren,flopb1,Victo-2014} and the importance of the energy quantization and a few-level systems has led to the study of quantum heat machines highlighting the essence of this effect\cite{PhysRevE.72.056110,PhysRevE.76.031105,Campisi2016,PhysRevLett.120.170601,Levy2018,PhysRevE.100.012123,PhysRevE.103.062109,e23050536,PhysRevE.104.014149}.

Recently, the size-invariant shape transformation (SIST) technique has been introduced to completely separate size and shape effects from each other\cite{aydin7}. By applying this technique on the domain that particles are confined, one can define control variables uniquely characterizing the shape of the domain. Then, for strongly confined systems, quantum shape effect is defined as the characteristic changes on physical properties due to changes in the corresponding shape variable controlling the shape. SIST lies in the heart of quantum shape effects. Quantum shape effects can cause classically impossible thermodynamic behaviors, most notably allowing spontaneous thermodynamic transitions to lower entropy states\cite{aydin7}. Among many possible applications, it is demonstrated that they can be used to design novel thermodynamic cycles\cite{aydin7} and to harvest energy based on core-shell nanostructures\cite{aydin11}. In Fermionic systems, Fermi level can effectively be controlled by shape to utilize the oscillations in thermodynamic and transport properties caused by quantum confinement\cite{aydin12}. All of these quantum shape-dependent behaviors can be traced back to their roots in the eigenspectra of the confining media. Both to develop a deeper understanding for quantum shape effects and to explain the fundamental underlying reasons of their consequent phenomena, examination of the eigenspectra under the SIST is needed.

\begin{figure*}[t]
\centering
\includegraphics[width=0.7\textwidth]{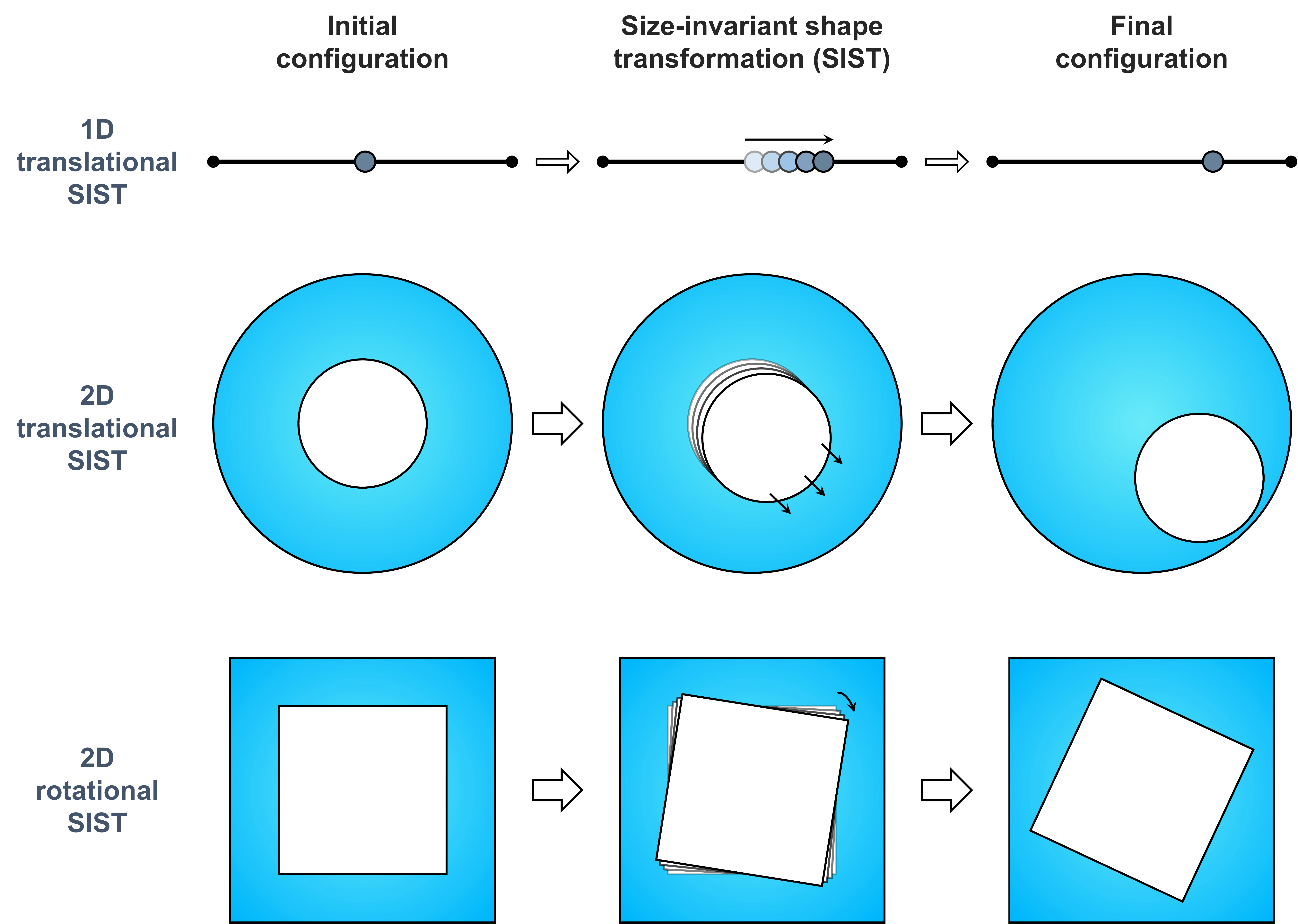}
\caption{Examples of size-invariant shape transformation (SIST). First and last columns show the initial and final configurations of the examined shapes, whereas the middle column show the applied transformations. First row shows the translation of an inner boundary inside a 1D domain. 2D version of the same concept is shown in the middle row, represented by nested circles. Rotation of the inner domain is another transformation in 2D. Examples can be extended to higher dimensions without loss of generality. All transformations preserve the sizes under Lebesgue measure while changing the shape of each domain.}
\label{fig:sist}
\end{figure*}

In this paper we start by exploring the properties of the Dirichlet spectrum of the Laplacian for confined geometries undergoing the SIST. We show that varying the shape of a domain by preserving its sizes amounts to creating a geometric coupling between otherwise independent levels, resulting a nonuniform energy level scaling that is characterized by two distinct changes in the Dirichlet spectrum: (1) The decrease of smallest eigenvalue, which we refer to as the ground state reduction. (2) Changing of the gaps between eigenvalues so that the formations of degeneracy or level splitting, depending on the initial configuration spectra and the type of the SIST that is applied. Moreover, we explain the underlying reasons for these changes. The first mechanism occurs as a result of the fact that shape of the local regions inside the domain becomes similar to the shape of an $n$-sphere, which affects the spectra in a way that reduces the first eigenvalue, as a consequence of Rayleigh–Faber–Krahn inequality. Apart from being a direct measure of sphericity, the radius of the inscribed $n$-sphere serves also as a good measure of the ground state reduction, which also helps to understand its underlying cause: increase of local breadths. The second mechanism in turn arises from the requirement that the asymptotic behavior of the eigenvalues must be the same under the effect of ground state reduction, as a consequence of the Weyl law. These two phenomena underlie the reason for the nonuniform level scaling and its counterintuitive thermodynamic consequences, such as spontaneous and simultaneous decrease in free energy and entropy. In addition, we provide a physical understanding for such behaviors by proposing the existence of a quantum thermodynamic phenomenon, directly coming from our spectral analysis, which we call quantum thermal avalanche. Our findings provide a fundamental understanding and explanation on the peculiarities observed in the thermodynamic properties of particles due to the quantum shape effect.

\section{Transforming the shape while preserving the size}

Size of a domain is characterized by geometric size parameters that are defined under Lebesgue measure as volume $\mathcal{V}$, surface area $\mathcal{A}$, peripheral length $\mathcal{P}$ and the number of vertices $\mathcal{N_V}$. There is a simple and continuous way of changing the shape of domain without altering any of these size parameters. It is called isometric (or more precisely size-invariant) shape transformation, which is visualized in Fig. 1 for various carefully chosen domains. The procedure of SIST can be described as follows: Consider any type of fixed domain shape in any dimension and introduce an inner domain with boundary into it. It can be a dot or a finite length in 1D, or any other shape in higher dimensions that can be removed from the considered domain. Removing the inner domain leaves the remaining domain that we are interested in, e.g. denoted by blue regions in Fig. 1. Now, one can apply translation or rotation to the inner domain that has been removed, which amounts to the continuous variation of the boundaries of the remaining domain. It is easy to notice that sizes of the domain stay fixed under these transformations. In other words, SIST is not only volume-preserving and area-preserving, but also a periphery-preserving and vertices-preserving operation. It preserves all geometric size variables together. We show only a few examples in Fig. 1, but actually this technique is quite general and can be applied to objects in any shape or dimension. It allows us to construct arbitrarily-shaped domains where we can change the shape without altering any of the geometric size parameters.

\begin{figure*}[t]
\centering
\includegraphics[width=0.95\textwidth]{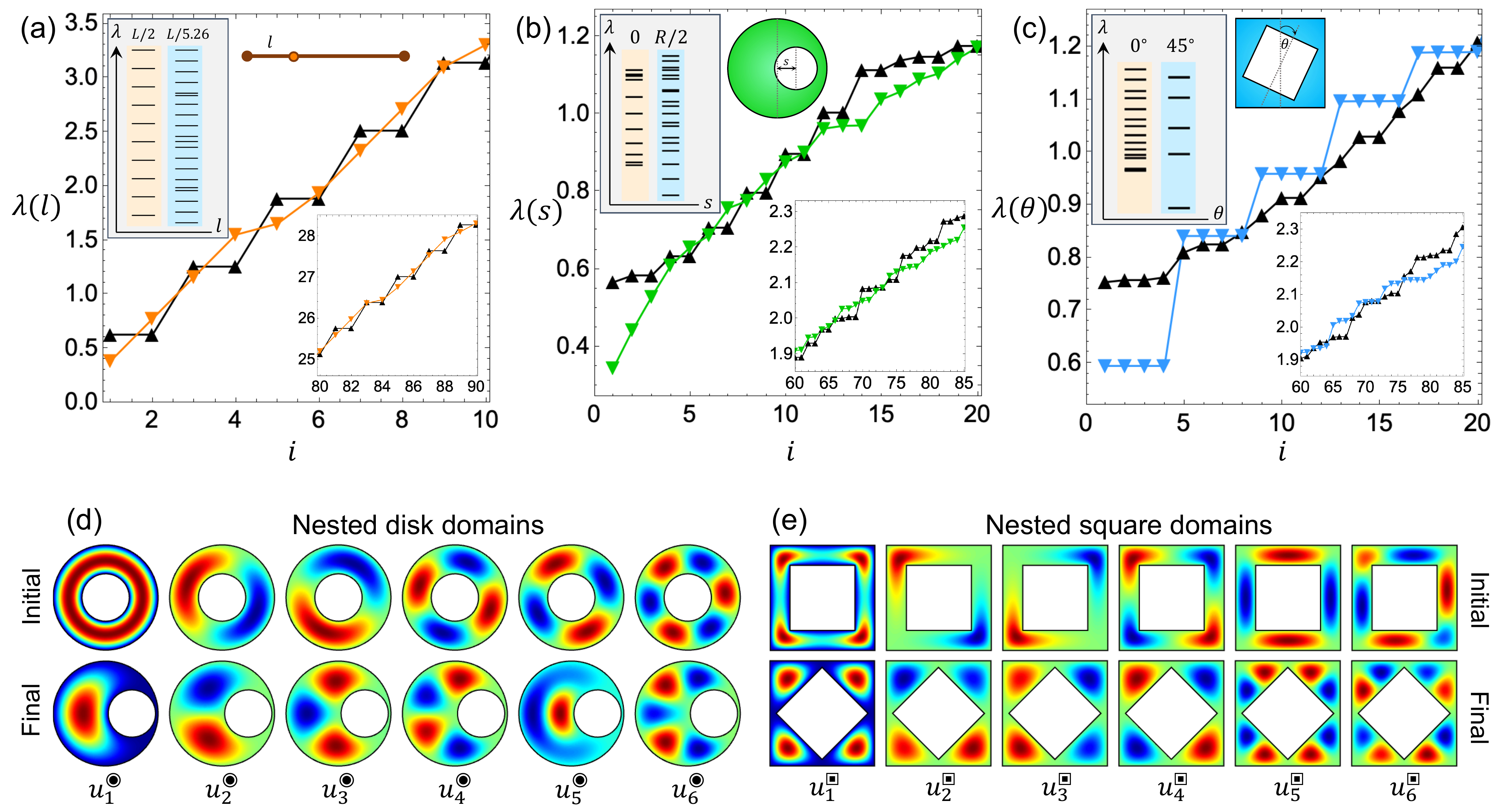}
\caption{Nonuniform level scaling in spectra under size-invariant shape transformations (SISTs) due to geometric level couplings. Eigenvalue $\lambda$ versus its number $i$ plots for (a) 1D domain with length $L$ undergoing translational SIST, (b) nested 2D disks with outer disk radius $R$ undergoing translational SIST and (c) nested 2D squares undergoing rotational SIST. SISTs are characterized by the shape variables $l, s, \theta$, respectively, whose meanings are illustrated in the inset figures showing the domains. Geometric level couplings are created by the corresponding shape variables. Eigenspectra of the initial (final) configurations are given by the black (colored) up(down)-triangles. For easy comparison, the first 20 eigenvalues are also shown as level-line plots in the boxed inset figures. Values of the shape variables for initial and final configurations are typed on top of their respective level spectra. An arbitrary portion of the higher eigenvalue spectra are also given as inset figures on the bottom-right of each subfigure. (d) First six eigenfunctions of nested disk and (e) nested square domains. Top and bottom rows show initial and final configurations respectively.}
\label{fig:eigs}
\end{figure*}

\section{Nonuniform level scaling due to geometric level coupling}

Consider the Dirichlet problem for the Laplacian on a domain $\Omega$
\begin{equation}
\begin{split}
\nabla^2 u+\lambda u=0 \;\;\;\text{on}\; \Omega, \quad u\vert_{\partial\Omega}=0,
\end{split}
\end{equation}
where $\lambda$ and $u$ are the Dirichlet eigenvalues and eigenfunctions and $\partial\Omega$ is the domain boundary. Eq. (1) is the Helmholtz equation and this eigenvalue problem is equivalent to finding fundamental and overtones of a drum\cite{kac66,Trefethen2006,RevModPhys.82.2213}, modes of a waveguide\cite{exner1,dcfib2,PhysRevE.74.016201,DellAntonio_2010,Delitsyn2012,PhysRevA.87.042519}, or a quantum billiard\cite{PhysRevLett.53.1515,PhysRevLett.100.204101,PhysRevE.101.032215} or finding the energy levels of electrons confined in a box with shape $\Omega$\cite{doi:10.1126/science.262.5131.218,mat2D}.

Now think of a drumhead with an elastic membrane in one of the shapes of 2D domains shown in Fig. 1. By applying SIST, we can prepare another drumhead and compare how they sound. Although the surface area $\mathcal{A}$ as well as all other size variables (such as the lower-dimensional ones, i.e. $\mathcal{P}$ and $\mathcal{N_V}$) of both drumheads are exactly equal to each other, spectra of the membranes will be different\cite{aydin7}. In other words, the domains constructed by SIST are not isospectral. They will sound different as the normal modes of the membranes would be different. In fact, under SIST, the most distinct one of all modes will be the fundamental mode of the membranes, which we will explore in next section. 

We consider three domains given in Fig. 1 and calculate each of their Dirichlet eigenvalues and eigenfunctions by numerically solving Eq. (1). We investigate the variations of their eigenspectra under their respective SISTs in Fig. 2. The first example we look at is a 1D domain with an inner boundary and the translation of this inner boundary to either left or right. One can thing of this as a particle in a box separated by a movable but impenetrable partition. This is actually the simplest case of a SIST which is comprehensively examined in Ref.\cite{origin}. The distance of the inner boundary (movable partition) from the outer one (the left boundary is chosen as a reference) is denoted by $l$, which comes out as the shape control variable of this 1D translational SIST, see Fig. 2(a). The partition could be chosen to have any finite thickness, here we choose it as zero thickness. For this simple system, eigenspectrum can be obtained analytically as the union of the eigenvalue spectra of left and right compartments,
\begin{equation}
\lambda_i(L,l)=\lambda_i^{\prime}(l)\cup \lambda_i^{\prime}(L,l)=\biggl\{\frac{\pi i^{\prime}}{l}\biggr\}\bigcup \biggl\{\frac{\pi i^{\prime}}{L-l}\biggr\},
\end{equation}
where $i,i^{\prime}=1,2,3,\ldots$ eigenstate numbers. We plot the first 10 eigenvalues in Fig. 2(a), eigenvalue number being denoted by $i$. The spectrum of the initial configuration (inner boundary is chosen at the center, $l=L/2$) is given by the black up-triangles and of the final configuration (inner boundary is at a close distance to the outer boundary, $l=L/5.26$) by the orange down-triangles. The values are presented as joined data for convenience. We plot first 20 eigenvalues also as a level-line plot in the boxed inset figure to make easier comparisons of spectral modifications. An arbitrarily chosen higher portion of the eigenvalue spectrum is plotted as another inset figure. Similar convention is followed for the other shapes that we consider. In Fig. 2(b) we examine the spectrum of the second domain composed of nested disks (smaller one is removed from the larger one) undergoing 2D translational SIST with a shape variable $s$ being the distance of inner disk's center from the center of outer disk. Initial configuration is chosen as concentric ($s=0$) and in the final configuration inner disk is slided to the right to be centered at $s=R/2$ where $R$ is the radius of the outer disk. As the third example, the spectra of nested concentric squares undergoing 2D rotational SIST is shown in Fig. 2(c). Here SIST is characterized by the rotation angle of the inner square, $\theta$, initially at $\theta=0^{\circ}$ and finally at $\theta=45^{\circ}$.

As is seen in Fig. 2, all types of SISTs cause nonuniform scaling of the eigenspectra. This is a peculiar behavior coming from the creation of a geometric coupling between the levels due to SISTs. Normally, eigenstates are influenced uniformly and linearly from the changes affecting the spectrum. For example, changing the length of a domain in any direction causes a uniform/linear scaling of the levels in that particular direction (e.g. doubling the length reduces the energy to a quarter in a particle in a box). Or, opening an external field shifts the spectrum as a whole, again corresponds to a uniform/linear scaling. In fact, all conventional thermodynamic control variables like volume, temperature, external field parameters, etc. cause linear transformations (uniform scalings or shifts) on the eigenspectra. On the other hand, the property of size-invariance induces a geometric coupling between otherwise independent levels. This can be most easily seen in Eq. (2) where the spectrum of the domain is composed of the union of the spectra of two spatially independent but geometrically coupled domains. The geometric coupling parameter is the position of the partition, $l$. Due to this additive coupling, levels are no longer independent and changes in $l$ nonuniformly affects the spectrum. The same phenomenon occurs in the other two type of SISTs. However, shape variables of the domains in higher dimensions have complicated functions of the eigenvalues and in general it is not possible to obtain the spectrum analytically to examine the exact functional forms of these geometric couplings. Nevertheless, even examining them numerically provides important insights on the nature of the phenomenon of nonuniform level scaling. 

While nonuniform level scaling provides a global perspective on the eigenspectra, here we go beyond and look further in detail to find out the underlying reasons for these type of modifications in the eigenspectra due to SISTs. There are two notable common modifications. The first and the most noticeable one is the lowering of the first eigenvalue, as it can be directly seen in all three cases that we consider in Fig. 2. The second is the breaking of degeneracies or equivalently the splitting of levels. Both have important consequences on the physical properties of particles confined in such systems. In the next section, we examine the underlying reason for the lowering of the first eigenvalue.

\section{Ground state reduction due to sphericity}

Before explaining the lowering of the first eigenvalue under SIST, it is tempting to question the direction of the transformation. After all we could reverse the initial and final configurations, which would still be considered as a SIST, but the first eigenvalue would have been increased in such a case. Essentially, there is a physical reason for choosing the direction of the transformation in the specified way. We choose the initial configurations in such way that it maximizes the Helmholtz free energy of the non-interacting particles confined inside the domains at constant temperature. As such, when systems are perturbed from their respective initial configurations, they spontaneously transition into their respective final configurations under quasi-static process. Since all thermodynamic state variables stay constant under this transition except the corresponding shape variable, the changes in free energy are solely due to the shape transformation.

\begin{figure*}[t]
\centering
\includegraphics[width=0.95\textwidth]{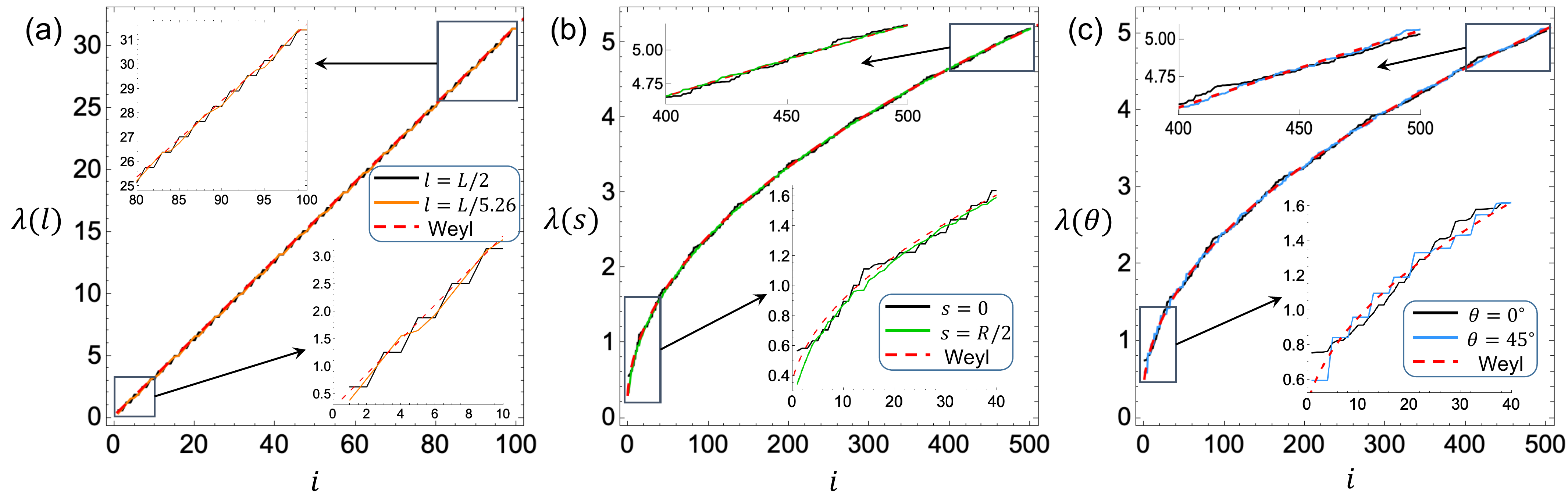}
\caption{Accuracy of Weyl law in predicting the asymptotic as well as average behaviors of eigenspectra of size-invariant domains having different shapes. Weyl law, however, fails to predict any shape difference. Eigenvalue spectra of (a) 1D domain under translational SIST, (b) 2D domain composed of nested disks under translational SIST and (c) 2D domain composed of nested squares under rotational SIST. Initial and final configurations are represented by black and colored curves respecively. Weyl curves are found simply by solving Eq. (4) for $\lambda$. Inset figures show the zoomed versions of the selected portions of spectra.}
\label{fig:weyl}
\end{figure*}

In quantum confined nanostructures, the sizes are comparable to the thermal de Broglie wavelength of particles so that only a few states can be thermally excited. Under such conditions, the influence of SIST on the physical properties of spatially confined particles is called quantum shape effect\cite{aydin7}. Quantum shape effects can be quantified analytically by invoking the quantum boundary layer concept\cite{qbl,uqbl,nanocav} and defining an effective $n$-volume considering the overlaps of quantum boundary layers\cite{aydin7, aydinphd}. See Appendix for details. Both the quantum boundary layers and their overlaps are temperature dependent quantities which brings both shape and additional temperature dependence to thermodynamic properties. Overlaps of quantum boundary layers occur when inner and outer boundaries come very close to each other. They increase the effectively available domain for particles to occupy. As long as the inner boundaries are at a far enough distance from the outer ones, overlaps won't be appreciable and quantum shape effects vanish. The more the overlap, the larger the quantum shape effects. Hence, we choose the direction of SIST in the way that quantum shape effects increase. Final configurations have more shape-dependence than the initial ones. Thus, reduction of the ground state happens to be a characteristic property of the quantum shape effect.

To understand how quantum shape effects lead to a persistent lowering of the ground states, we start by noticing the immediate connection between the eigenspectrum and the geometric size parameters. This connection is quantified by Weyl law describing the asymptotic behavior of the Dirichlet eigenvalues of the Laplacian in terms of the geometric size parameters under the Lebesgue measure\cite{weyl11,baltes,weylcomput20}. The asymptotic expression of the number of eigenvalues less than $\lambda$ for the Helmholtz equation can be written in its general, $D$-dimensional, closed form \cite{aydinphd} as,
\begin{equation}
W_D(\lambda)= \sum_{n=0}^D\left(-\frac{1}{4}\right)^{D-n}\left(\frac{\lambda}{2\sqrt{\pi}}\right)^n\frac{\mathcal{V}_n}{\Gamma\left[n/2+1\right]},
\end{equation}
where $\mathcal{V}_n$ is the $n$-dimensional volume under the Lebesgue measure ($\mathcal{V}_3\rightarrow \mathcal{V}$, $\mathcal{V}_2\rightarrow \mathcal{A}$, $\mathcal{V}_1\rightarrow \mathcal{P}$, $\mathcal{V}_0\rightarrow \mathcal{N_V}$). Since we are dealing with 1D and 2D domains, let's write their explicit Weyl expressions (we also write 3D one for completeness) respectively as
\begin{subequations}
\begin{align}
W_1(\lambda)= & \frac{\mathcal{P}}{\pi}\lambda-\frac{\mathcal{N_V}}{4}, \\
W_2(\lambda)= & \frac{\mathcal{A}}{4\pi}\lambda^2-\frac{\mathcal{P}}{4\pi}\lambda+\frac{\mathcal{N_V}}{16}, \\
W_3(\lambda)= & \frac{\mathcal{V}}{6\pi^2}\lambda^3-\frac{\mathcal{A}}{16\pi}\lambda^2+\frac{\mathcal{P}}{16\pi}\lambda-\frac{\mathcal{N_V}}{64}.
\end{align}
\end{subequations}

\begin{figure*}[t]
\centering
\includegraphics[width=0.95\textwidth]{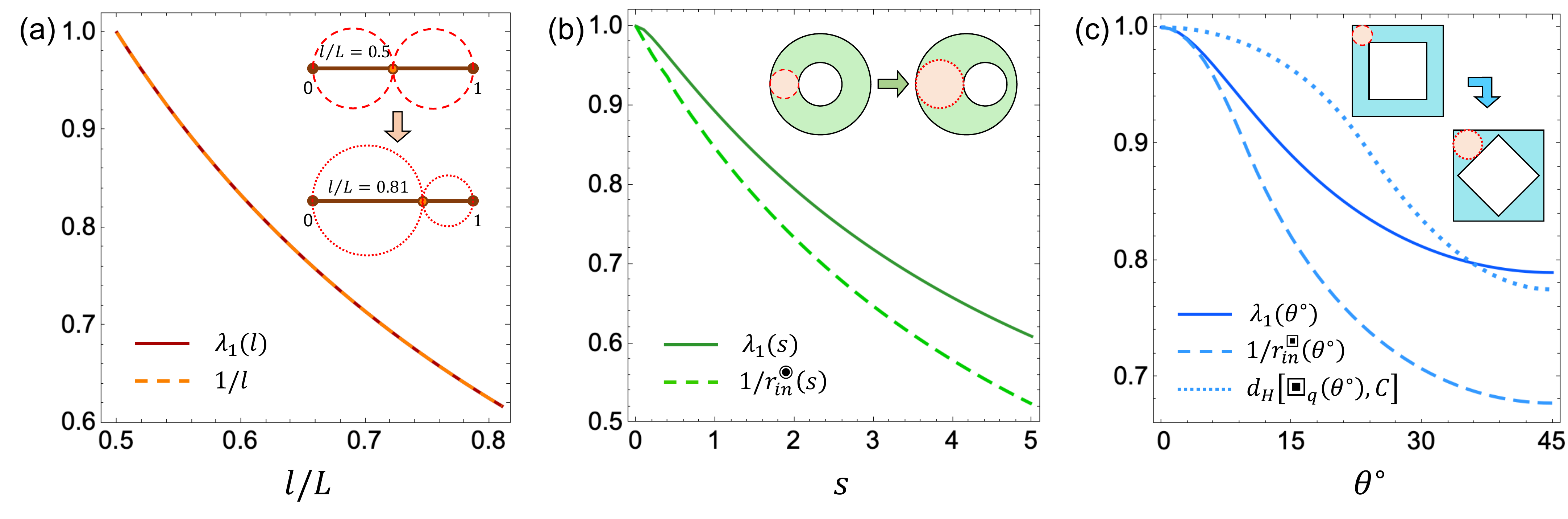}
\caption{Decrease of the ground state with respect to shape variables can be understood by the increase of breadth and sphericity of the local parts of the domain. All values are normalized by the values of their respective initial configurations. Solid curves represent the variation of the first eigenvalue with the corresponding shape variable of (a) 1D domain with an internal boundary (b) 2D domain of nested disks and (c) 2D domain of nested squares. Local breadth is measured by the radii of inscribed circles, which are shown by dashed curves. Ground state reduction is associated with the enlargement of the inscribed circles. As local breadths and first eigenvalues are inversely proportional, inverses of the local breadth measures are plotted. Dotted function represents the Hausdorff distance measuring the sphericity. The shorter the Hausdorff distance, the more sphere-like the object is.}
\label{fig:measures}
\end{figure*}

In Fig. 3, we compare the actual eigenspectra of initial and final configurations (i.e. the ones that are the most radically different) under SIST with the predictions of Weyl law for different domains. Weyl law quite accurately predicts the behaviors of eigenvalues even for low-lying ones by passing around their averages\cite{aydin3}. It is also appropriate to mention here that, the fact that Weyl law can successfully produce all the lower dimensional quantum size effect corrections\cite{aydin3,aydin8} relies on its remarkable success in predicting the behaviors of eigenvalues. This makes it a reliable tool to analytically study quantum size effects in confined systems, via for example Weyl density of states\cite{PhysRevA.87.042519,aydin3,aydin8,origin}. 

Dirichlet eigenvalues and eigenfunctions are directly determined by the shape of the domain. Besides the global size and shape of domains, local shape information is also embedded in the eigenspectra\cite{Benguria2012,shape2Ddirichlet,KHABOU2007141}. In fact, for concave or hollow domains (like the ones we consider here), the shapes of the local parts of the domain could be even more decisive in determining the characteristic spectra\cite{aydin7}. All geometric size parameters remain unchanged under SIST, which means Weyl law gives exactly the same values and cannot distinguish the shape difference, see Fig. 3. Since Weyl law is basically an asymptotic theorem, it cannot predict the changes due to SIST. Nevertheless, it is possible to extract some useful information by studying the form of the Weyl expressions. For instance, Weyl law states that the bulk terms ($\mathcal{P}$ in 1D and $\mathcal{A}$ in 2D) have the largest effect on the eigenspectrum. Although geometric size variables are preserved globally under SIST, arbitrarily selected parts of the domains can have different sizes locally. Therefore, we could get insights on the behavior of the first eigenvalue by investigating the local confinements, whose information is embedded in the eigenvalues and eigenfunctions. See for instance Fig. 2(d) and (e), where we plot the first six eigenfunctions for the initial and final configurations of nested disk and square domains respectively. Now imagine cutting the nested disk domain vertically into two halves and consider the left half during SIST (see how the first eigenfunction varies in Fig. 2(d)). We can notice that the first eigenfunction experiences as if the domain is larger in the final configuration compared to the initial one. Even though area stays fixed \textit{globally}, translating the inner disk to the right leaves more area \textit{locally} in the left part of the domain. The ground state eigenfunction cannot fit into the narrow parts of final configuration and experiences a larger domain overall, compared to the initial configuration. Then, from the inverse proportionality of size variables with eigenvalues, we expect the final configuration to have a lower first eigenvalue.

The similar argument can be made also for the 1D domain as well as 2D nested square domain. In 1D case, there is a larger length for the ground state eigenfunction to occupy when the inner boundary is closer to the outer boundaries compared to the initial configuration where the boundary is at the center. While the confinement of one part causing the deconfinement of the other during the movement of the partition, the overall effect is actually asymmetric because of the decay of occupation probabilities in the smaller part. Hence, when the boundaries come very close to each other, they become almost indistinguishable (i.e. acting as if it is a single boundary) for the confined particles, which amounts to the local breadth of the larger part. 

Note that in the case of the nested square domain, the geometric size variables are the same even for each quadrant under SIST (quadrant is defined considering nested squares being concentric at the origin). Still, a triangular-like shape (quadrant at $\theta=45^{\circ}$) is perceived less confined for the ground state eigenfunction compared to a bended tube shape (quadrant at $\theta=0^{\circ}$).

To refer these less confined regions of a domain, we will use the word "breadth", whose third meaning in dictionary.com\cite{breadth} stated as "freedom from narrowness or restraint" describes quite well what we would like to verbalize. We argue that there is a direct correspondence between the lowering of the first Dirichlet eigenvalue and increasing of the local breadths of our domains. To quantify this, we propose the radius of the inscribed $n$-sphere (circle in 2D) as a measure. Inscribed circle is the largest circle that can fit into a 2D domain. Then, we conjecture that the larger the radius of the inscribed circle that we can fit into the domain, the lower the first eigenvalue:
\begin{equation}
\lambda_1\propto\frac{1}{r_{in}^{\Omega}},
\end{equation}
where $\lambda_1$ is the non-zero first eigenvalue and $r_{in}^{\Omega}$ is the radius of an inscribed circle of a domain $\Omega$.

The connection between the inverse radii of the inscribed circles of domains and their lowest Dirichlet eigenvalues is shown in Fig. 4. There we show the variation of the first eigenvalue (solid curves) with corresponding shape parameters and the predictions of our local breadth measure (dashed curves). For convenience of comparison, all curves are normalized to their initial values at their initial configurations. In 1D domain, the local breadth can directly be quantified by the inverse distance of the inner boundary to the outer one ($1/l$), which is equivalent to the inverse diameter of the inscribed circle in 1D case, Fig. 4(a). While local breadth measure exactly matches with the behavior of the first eigenvalue in 1D, this is trivial because the eigenvalue spectrum is in fact already inversely proportional to the domain lengths, e.i. the first eigenvalue is $\lambda_1(L,l)=\min(\frac{\pi}{l},\frac{\pi}{L-l})$.

2D analogue of the increase in local breadth due to translation can be seen in Fig. 4(b). In this case, the inner disk shifts to the right and increases local breadth in the left portion of the domain. Inverse of the increase of the inscribed circle radius matches quite well with the decrease of the first eigenvalue. The case of 2D rotational SIST is shown in Fig. 4(c), where again the inscribed circle method captures the functional behavior of the first eigenvalue under shape variation. As is seen, inscribed $n$-sphere method provides a good measure for capturing the inverse proportionality between the local breadths and the first Dirichlet eigenvalues.

A quite related concept here is the isoperimetric inequality which states that among all shapes the circle (or $n$-sphere in general) has the largest surface area for fixed peripheral lengths\cite{specbook2}. Immediate connection of this fact to the eigenspectrum is provided by Rayleigh–Faber–Krahn inequality, which states that among all shapes having equal area, the circle has the lowest first Dirichlet eigenvalue\cite{specbook2,Benguria2012,spechenrot,Henrot2018}. This suggests that between shapes of equal sizes, the one that resembles most to a circle should have the smallest first eigenvalue. In fact in literature, inscribed circle has already been used as one of the methods that are proposed to distinguish shapes from each other and quantify their resemblance\cite{surveyshape,ZHENG2020122}. 

Another measure for local breadths is the sphericity of a domain. Sphericity is defined as a measure of how spherical (circular in 2D) an object is. There are several measures of sphericity in the literature and one of them is in fact the ratio of the radii of circumscribed and inscribed circles\cite{spherifirst,ZHENG2020122}. This is exactly the same as our local breadth measure since the radii of circumscribed circles do not change under SIST for any domain shape. In fact, isoperimetric inequality already implies that the local breadth and sphericity carry the very same meaning. The fact that circle has the largest area of all isoperimetric objects, makes it the best choice to describe the breadth of a domain which quantifies the largeness of local areas.

Hausdorff distance is one of the reliable measures for the similarity of two domains\cite{varan,VANKREVELD2022101817,BASRI19982365,shapesim}, which is also used in symmetrization methods\cite{geoin}. Hausdorff distance representing the similarity of the shape of a domain $\Omega$ to the shape of a disk $C$ is defined \cite{varan} as
\begin{equation}
d_H(\Omega,C)=\max\left[\sup_{\omega\in\Omega} \inf_{c\in C}d(\omega,c),\sup_{c\in C} \inf_{\omega\in\Omega}d(\omega,c)\right],
\end{equation}
where $d(\omega,c)$ is the distance between the point $\omega$ in the region $\Omega$ and point $c$ in the region $C$. It gives best results for non-hollow domains, so we can use it to quantify the similarity of the quadrant in nested square domains to a circle. Consider dividing the nested square domain into its quadrants by taking the origin at the center of the squares. Each quadrant is identical to others as they are symmetric about the axes at all rotational configurations. At the initial configuration of $\theta=0^{\circ}$ it starts from a bended tube shape and transitions into an obtuse triangle shape in the final configuration of $\theta=45^{\circ}$. One can intuitively say that the final configuration looks more like a circle. When we measure the Hausdorff distance of the formed shapes to a disk, we find this is indeed the case. The shorter the Hausdorff distance, the more spherical the object is. Since, sphericity is related to the first eigenvalue via Rayleigh–Faber–Krahn inequality, we conjecture that Hausdorff distance to a disk should also give an estimate of the behavior of the first eigenvalue such that
\begin{equation}
\lambda_1\propto d_H(\Omega_p,C),
\end{equation}
where $\Omega_p$ with the subscript $p$ denoting a chosen part (one quadrant in this case) of the domain $\Omega$. In Fig. 4(c), the variation of Hausdorff distance with the rotation angle is plotted by dotted curve, which accurately describes the behavior of the first eigenvalue. Note that both Hausdorff and incircle methods are proposed just to understand the ground state reduction qualitatively. One should not expect a perfect quantitative match, as the actual behaviors of spectra are determined by the exact domain boundaries. However, we demonstrate that ground state reduction is a consequence of the increased breadth and sphericity of the local parts of the domain, which appears as one of the characteristic properties of SIST.

\section{Level splitting and degeneracy due to size-invariance}

\begin{figure*}[t]
\centering
\includegraphics[width=0.95\textwidth]{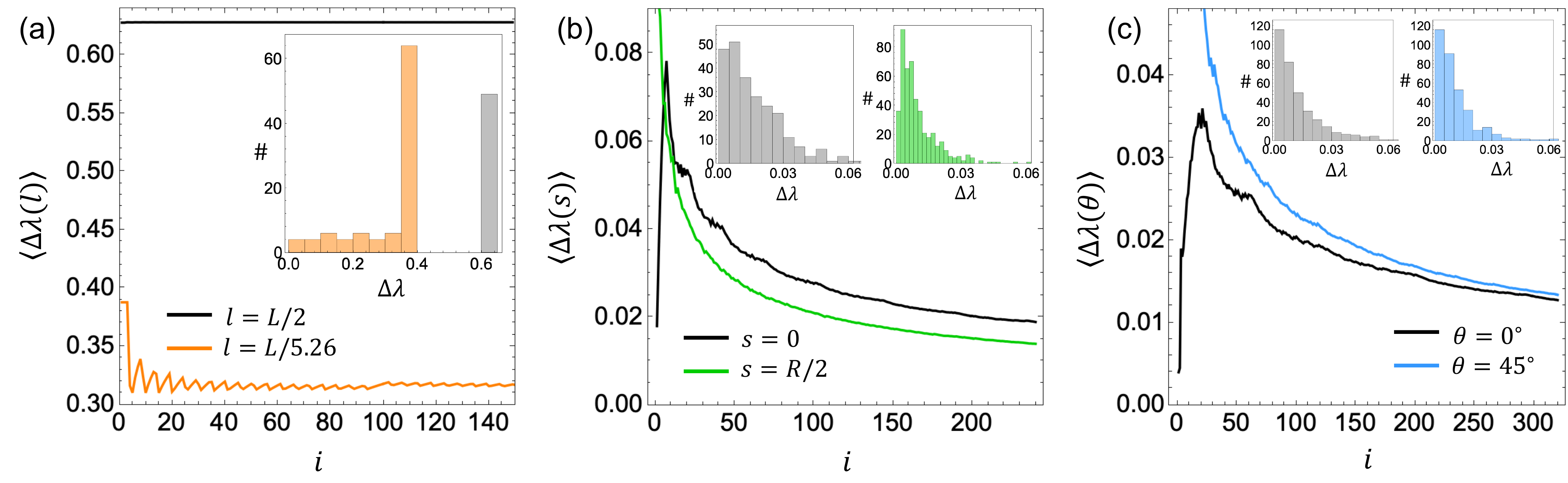}
\caption{Mean level spacings. Variation of the average differences (without degeneracies) between the levels (eigenvalues) with increasing level number $i$ for (a) 1D domain, (b) 2D nested disks and (c) 2D nested squares. Black and colored curves represent the initial and final configurations respectively. Inset figures show the histogram of the level differences. Gray and colored histograms represent the initial and final configurations respectively.}
\label{fig:mean}
\end{figure*}

The other characteristic change in the spectra of domains under SIST is the changes in spectral gaps. Level-line figures are shown in the insets of Fig. 2(a,b,c). While the ground state reduction was a common characteristic of any SIST, changes in spectral gaps are different for translational and rotational SISTs. In the case of translational SISTs, the initial configurations are the most symmetric ones. As it can be seen from Fig. 2(a) and (b), degeneracies are prevalent at the initial configurations, but they vanish under translational SISTs, which breaks the symmetry inside the domain and cause the level splitting. In such domains, breaking of degeneracy and occurrence of level splitting as a follow-up can be understood by the breaking of translational symmetry. On the other hand, under rotational SIST an opposite behavior occurs. More degeneracies are formed when the nested square domain undergoes rotational SIST, see Fig. 2(c). Essentially, this is intimately related with the characteristic shape of the domains, however, the main difference compared to the translational case is that rotational SIST preserves the axial symmetries of the initial configuration. Both initial and final configurations incorporate degeneracies for nested square domains. These degeneracies are visualized in eigenfunctions which can be seen in Fig. 2(e). The biggest difference is that there are four almost distinct chambers forming in each quadrant of the the final configuration. Since the apex length (closest point of inner and outer boundaries) is chosen to be very narrow, low-lying eigenfunctions could not see other chambers and nest themselves into almost distinct chambers, promoting the four-fold degeneracy. In the initial configuration, on the other hand, the apex length (basically corresponds to the tube width) is large enough for low-lying eigenfunctions to fit, which creates less of a degeneracy compared to the final configuration. That's why eigenvalues are closer to each other at $\theta=0^{\circ}$ and distant at $\theta=45^{\circ}$ configurations.

We can quantify the changes in the spectral gaps by studying the level statistics. In Fig. 5, we plot the variation of the averaged differences between successive eigenvalues by removing degeneracies. Black and colored curves represent the initial and final configurations respectively. Mean spacing is defined by
\begin{equation}
\langle\Delta\lambda\rangle_N=\sum_{i=1}^N\frac{\lambda_{i+1}^{\prime}-\lambda_{i}^{\prime}}{N},
\end{equation}
where prime superscripts denote the non-degenerate eigenvalues and $N$ is the total number of eigenvalues that are considered. The degeneracies are removed in order to prevent them improperly lowering the average values, as we are investigating specifically the gaps in this case. As we expected, for translational SISTs spectral gaps are larger in the initial configurations compared to the final ones, see Fig. 5(a) and (b) where black curves stay on top of the colored ones. On the other hand, the behavior is reversed and black curve stays consistently below the blue curve in Fig. 5(c) showing that spectral gaps are larger in the final configuration compared to the initial one in rotational SIST. This is also pretty clear in Fig. 2(c). 

We also plot histograms of non-degenerate level spacings as inset figures in Fig. 5. In the 1D case, spacing distributions are too wide apart to assign any statistical behavior. Initial configuration has a constant level spacing and final configuration has a few spacing values, Fig. 5(a). In the nested disk case, when the system is perturbed from the initial configuration the nearest-neighbor level spacings transition to be of the Wigner-Dyson type suggesting a chaotic behavior, see green histogram in Fig. 5(b). In the nested square case, the level spacings unambiguously exhibit Poisson type statistics suggesting a regular behavior\cite{qchaos}. 

There is an immediate connection between the ground state reduction and the behaviors of the spectral gaps. When a domain undergoes a SIST, the first eigenvalue decreases. However, as a result of size-invariance condition, the domain also has to keep the same asymptotic behavior in accordance with the Weyl law. The only way to preserve the asymptotic behavior is to compensate the ground state reduction by opening spectral gaps in the excited states. This is clearly observed in Fig. 2(c) and it occurs also to some extent in nested disk domains as near-ground state energy spacings could increase (e.g. the low-lying eigenvalue behavior in Fig. 5(b)). It eventually leads to a behavior where both eigenspectra fluctuates around their Weyl asymptote. Since low-lying states are prominent in determining the physical properties of confined systems, the compensation of the ground state reduction by the gap openings serves as an important characteristic of the spectra of systems exhibiting the quantum shape effect.

\begin{figure*}[t]
\centering
\includegraphics[width=0.95\textwidth]{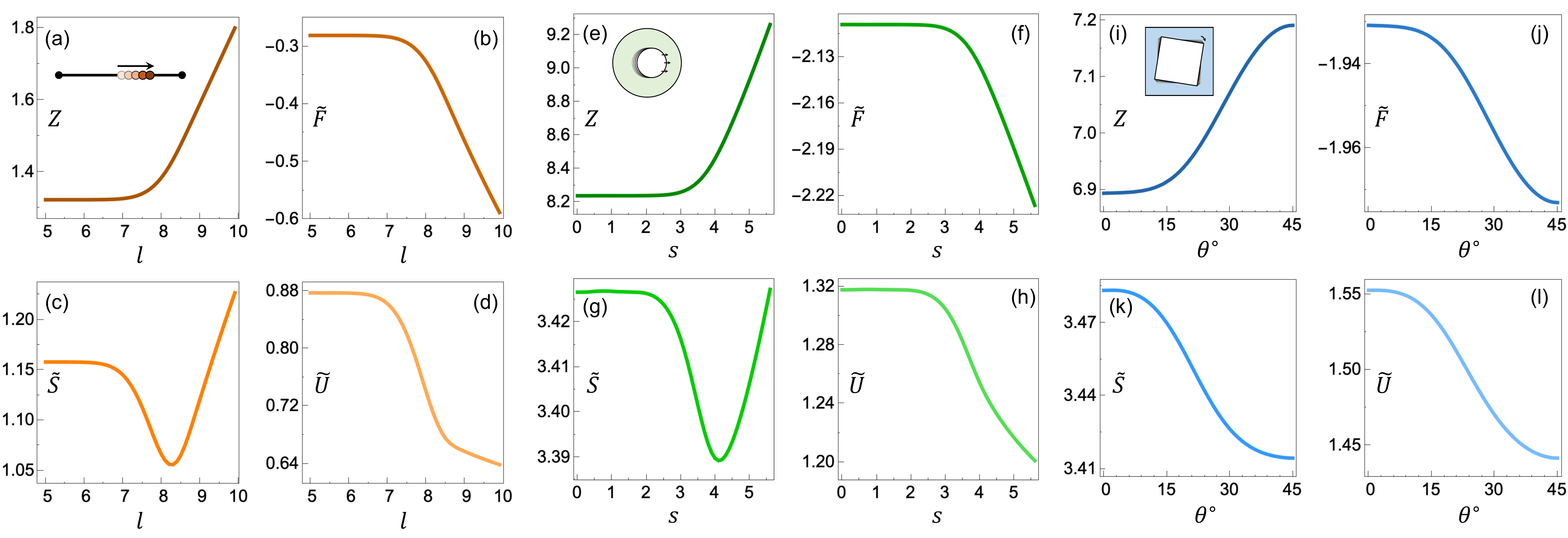}
\caption{Variation of thermodynamic properties with the corresponding shape variables. Curves with the shades of red, green and blue represent 1D domain with moving partition, 2D nested domain with translating disk and 2D nested domain with rotating square shown as inset respectively. Thermodynamic properties are shown from darker to lighter shades of colors as partition function, free energy, entropy and internal energy, where the last three are normalized (to $k_BT$ and entropy to $k_B$).}
\label{fig:thermo}
\end{figure*}

\section{Weakly size-invariant anisometry transformation}

Before going into the physical consequences of the spectral changes due to SIST, we will also explore a subordinate effect of it. Rather than keeping all of the size parameters constant, one can consider keeping the bulk parameter (area in 2D) fixed and changing the anisometry of the domain. Take for instance a square domain and turn it into a rectangle by changing the side lengths in a way to keep the area fixed. Although such transformation is area-preserving, it cannot preserve the periphery at the same time. We can call such a transformation a weakly size-invariant anisometry transformation. Here the term "weakly" indicates that it is not fully size-invariant but invariant with respect to only some of the geometric size parameters. 

Nevertheless, because of the area-preserving nature of the system, nonuniform level scaling in the eigenspectra for such weakly size-invariant anisometry transformations has already been observed \cite{PhysRevLett.120.170601,Levy2018,PhysRevA.99.022129,PhysRevE.104.044133}. As we argue here, such behaviors originate from the increased local breadth and obedience to the Weyl law. Sphericity of the square is higher compared to a rectangle with same area, making the square configuration having the lowest eigenvalue of all other anisometric rectangles. Also, weak size-invariance via area-preserving transformation causes splitting of the levels. Under any type of size-invariant transformation, the geometric information of the perpendicular directions are no longer independent. The condition of preserving sizes induces a geometric coupling between the initial and final configurations of the transformed geometry. In this weakly size-invariant anisometric transformation, a multiplicative coupling is created between the sizes of two orthogonal directions, e.g. the initial configuration is a square with the side length of $a$ and the final configuration is a rectangle with side lengths of $a^2/b$ and $b$. The eigenvalue spectrum of such a system becomes
\begin{equation}
\lambda_{i_1,i_2}(a,b)=\sqrt{\left(\frac{\pi i_1}{a^2/b}\right)^2+\left(\frac{\pi i_2}{b}\right)^2},
\end{equation}
where $i_1,i_2=1,2,3,\ldots$ are the eigenstate numbers of two orthogonal directions. Here, $b$ serves as the coupling variable mediating the geometric coupling between two orthogonal directions. When $a=b$, the square configuration is recovered. Hence, the variations with respect to the variable $b$ induce nonuniform scaling in the spectra. One drawback of this type of anisometry transformations is it is not quite easy to see how they can be achieved continuously under a quasi-static process. Continuous nature of the SIST processes makes them easier to physically realize. We won't go into the details of weakly size-invariant anisometry transformation, to keep the focus on the full size-invariance in this work.

\section{Peculiar thermodynamic behaviors: Entropy anomalies}

In this section, we discuss a physical example where these spectral modifications manifest themselves causing some unexpected physical behaviors. Consider a quantum particle confined in a box at thermal equilibrium at a temperature $T$ with a heat bath. The wave nature of the particle becomes significant when the boundaries of the domain are close enough to each other so that the sizes are in the order of thermal de Broglie wavelength of the particle. One cannot consider it as a point particle and continuous spectrum approximation becomes invalid. Hence, there are deviations from the ideal gas behavior due to quantum size and shape effects\cite{pathria,dai1,Dai_2004}. Under quantum shape effects, in particular, thermodynamic state functions exhibit peculiar behaviors never seen before in the classical thermodynamics of gases\cite{aydin7,aydinphd,aydin10,aydin11,aydin12}. 

Canonical partition function of a particle confined in a domain is written as
\begin{equation}
Z=\sum_{i}\exp\left(-\frac{E_i}{k_BT}\right),
\end{equation}
where $k_B$ is Boltzmann constant, $T$ is temperature and translational energies $E_i=\hbar^2k_i^2/(2m)$ with $\hbar$ being the Planck constant, $m$ being the particle mass and $k_i$ are the momentum eigenvalues (the same as $\lambda_i$) that depend on the corresponding shape variable for each case that is considered, i.e. $k_i(l)$, $k_i(s)$, and $k_i(\theta)$, respectively for 1D domain with partition position $l$, 2D domain with the inner disk position $s$, and 2D domain with the inner square rotation angle $\theta$. Let's assume a single particle for convenience, since the functional behaviors of the statistical thermodynamic quantities would be exactly the same as the many particle case because they are assumed to be non-interacting. Thermal occupation probability of a level $i$ is given by
\begin{equation}
p_i=\frac{1}{Z}\exp\left(-\frac{E_i}{k_BT}\right).
\end{equation}
Then, Helmholtz free energy, Gibbs entropy and internal energy are written respectively as
\begin{subequations}
\begin{align}
F=&-k_BT\ln Z, \\
S=&-k_B \sum_i p_i\ln p_i, \\
U=&k_BT\sum_i p_i E_i.
\end{align}
\end{subequations}

\begin{figure}[t]
\centering
\includegraphics[width=0.48\textwidth]{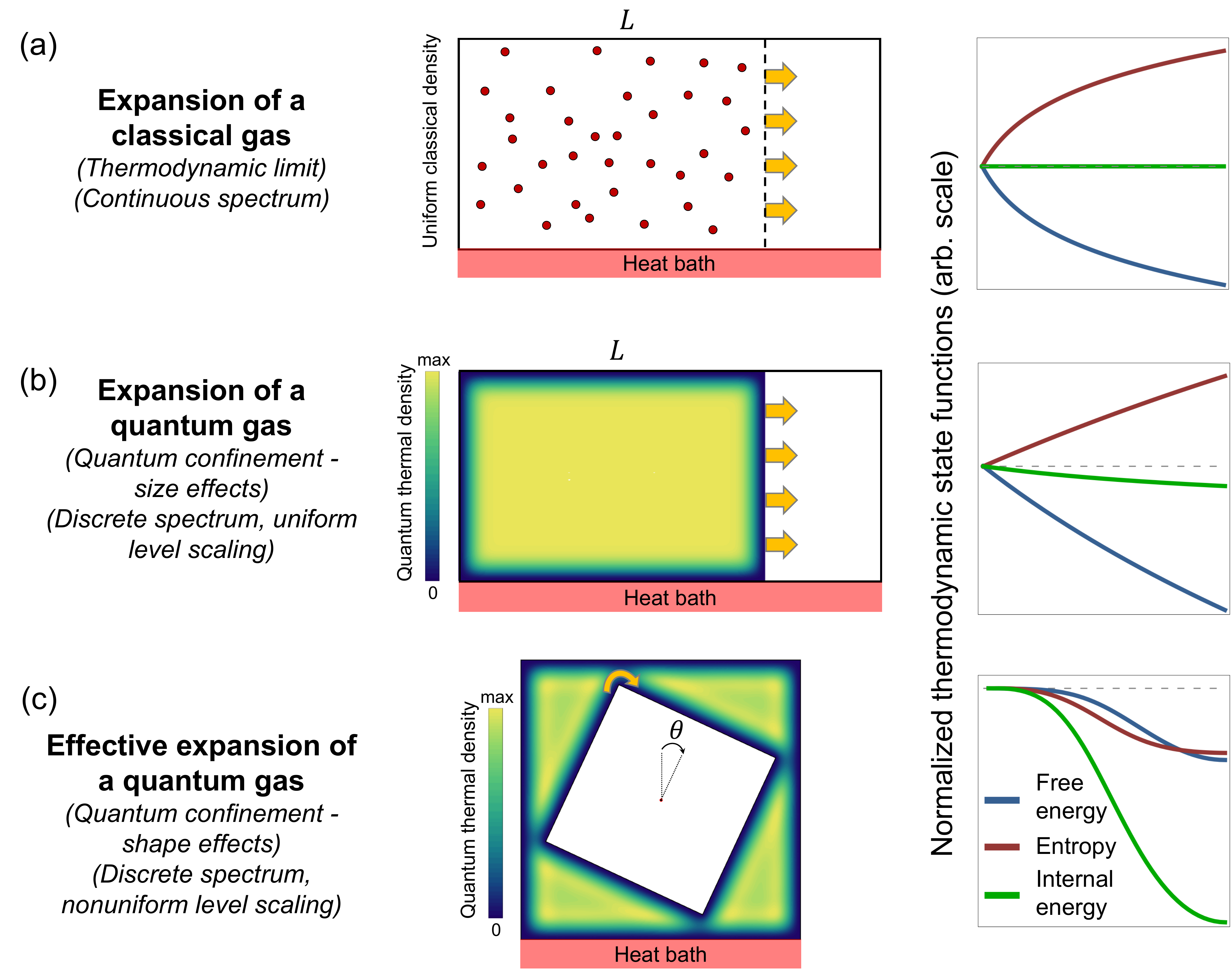}
\caption{Comparison of the isothermal expansions of unconfined (a) and confined gases with uniform (b) and nonuniform (c) level scalings. Blue, red and green curves represent the variations of free energy, entropy and internal energy respectively with the corresponding expansion parameters. Thermodynamic state functions are normalized to their initial values and given in arbitrary scale since we only emphasize their functional behaviors here.}
\label{fig:compare}
\end{figure}

In Fig. 6 we plot the functional behaviors of thermodynamic properties changing with the corresponding shape variables. Red, green and blue curves respectively represent the thermodynamic properties of 1D domain with moving partition, 2D nested domain with translating disk and 2D nested domain with rotating square. Partition function, free energy, entropy and internal energy are represented by the colors from darker to lighter shades respectively. In all cases the partition function increases, as the total occupation probabilities increase. The reason of this can be understood via the following analysis. Classically, partition function is a function of volume and temperature, as $Z\propto \mathcal{V}_n$ and $Z\propto T^{n/2}$. Therefore, classically one should not expect any variation in partition function under SIST, as the volume (as well as all other size parameters) stays constant. The changes in partition function can be understood by introducing the effective volume concept of the quantum boundary layer approach\cite{qbl,origin}. Although the apparent volume $\mathcal{V}$ of the object remains unchanged, the effective volume $\mathcal{V_{\mli{eff}}}$, which depends on the corresponding shape variable, changes. Some more information about the effective volume and quantum boundary layer concepts can be found in the Appendix. Effective volume is defined with additional terms including more complicated temperature dependencies. By definition, effective volume is also directly proportional with the partition function and they share the same functional dependency. Hence, the partition function mimics the behavior of effective volume which is increased due to the emergent overlaps of quantum boundary layers\cite{aydin7}. Due to these overlaps the domain becomes less confined locally (i.e. increase of local breadth). The partition function makes a slow start because the inner and outer boundaries are not close enough to each other for their quantum boundary layers to overlap (here we are not necessarily considering the zeroth order overlap, see for instance Fig. 10 in Ref.\cite{aydin7}). Once they are close enough, overlaps cause effective volume to increase exponentially (due to the enhanced ground state occupation), hence the partition function as well, Fig. 6(a). However, after some point the boundaries are so close to each other that effectively no particle can occupy the space in the squeezed part of the domain. Then, moving the inner boundary becomes no different than extending the domain length as a whole (the particle cannot distinguish the inner boundary from the outer one as we also mentioned in Sec. IV). In such a case, the usual volume (length in 1D) dependence of the partition function is recovered and it increases linearly with $l$, Fig. 6(a).

These characteristic behaviors of the partition function are directly reflected in other thermodynamic properties as well, shown in Fig. 6(b), (c) and (d). Thermodynamic properties change negligibly at the region where the partition function does not change much. In the exponential increase region, free energy, entropy and internal energy decrease together with the shape variable, which is a peculiar behavior having many counterintuitive consequences\cite{aydin7,aydinphd,aydin10,aydin11,aydin12}. In the linear increase region, free energy and internal energy decrease whereas entropy increase. This last behavior is ordinary as it is consistent with the expectations of classical thermodynamics. Characteristic functional behaviors of thermodynamic properties are quite similar in 1D and 2D translational SIST, compare red and green curves in Fig. 6. This is expected, since both operations are essentially the same (increasing the local breadth by squeezing the other parts of the domain), just realized in different dimensions for different shapes. In the 2D rotational SIST, on the other hand, we see more of a sigmoid-like increase in partition function. This leads to the simultaneous decrease of free energy, entropy and internal energy with the shape variable in their full ranges. Monotonic decreases in free energy and internal energy are expected, since the domain is effectively expanding under the rotational SIST, which is evident from the behavior of the partition function. This leads to a decrease in confinement energy which reduces the internal energy and effective expansion of the particles with increased $\mathcal{V_{\mli{eff}}}$ is accompanied by the decrease in free energy. Such thermodynamic behaviors have been shown to be important to precisely calculate the heat and work exchanges in quantum Szilard engines\cite{aydin10}.

\begin{figure*}[t]
\centering
\includegraphics[width=0.99\textwidth]{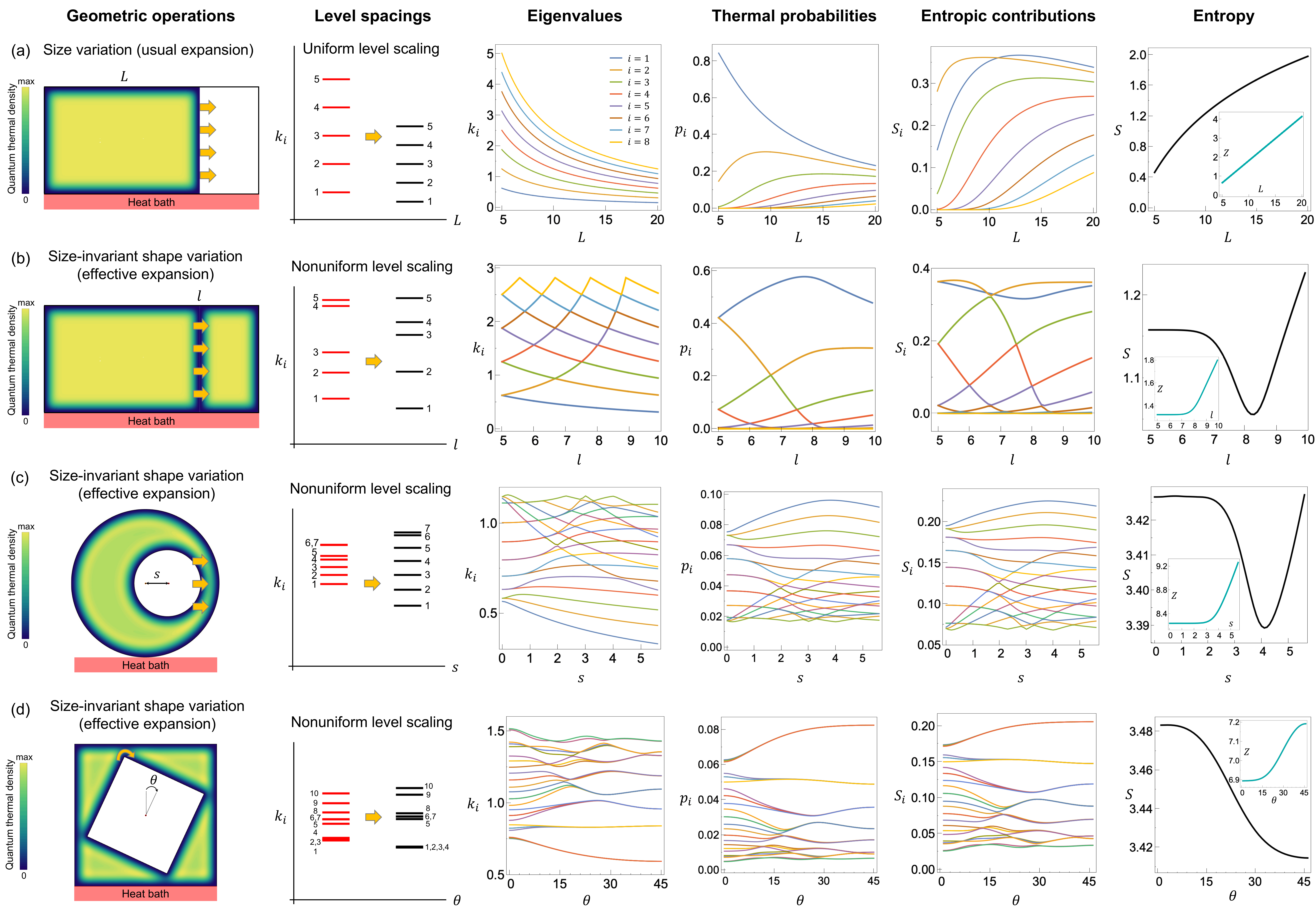}
\caption{Spectral and thermal signatures of entropy anomalies and consequences of nonuniform level scaling. Rows show the systems that are considered. Geometric operations and quantum thermal densities are shown for each system. Due to quantum confinement, density distributions are nonuniform even at thermal equilibrium. (a) Isothermal expansion of a quantum gas. The volume (length in 1D) of the gas increases as a whole. (b) Isothermal variation of the inner boundary (partition) of a box with a quantum gas. The actual volume stays constant, whereas effective volume of the gas increases. (c) Isothermal variation of the inner boundary in the nested disk domain. (d) Isothermal variation of the inner boundary in the nested square domain. Columns in each row represent respectively: the system, scaling of energy levels, variation of first few eigenvalues with respect to the corresponding size or shape variable, variation of thermal occupation probabilities of each state $i$, variation in the contributions of each state $i$ to the entropy and finally the total entropy of the system. Inset figures show the changes in the partition functions (corresponding to the expansion or effective expansions).}
\label{fig:entropy}
\end{figure*}

The most interesting one of the shape-dependent thermodynamic behaviors is the spontaneous and simultaneous decrease of free energy and entropy under an isothermal, quasi-static SIST process\cite{aydin7}. Decrease in entropy is much more difficult to understand solely from the effective volume perspective. To understand the peculiarity of such a decrease, consider an isothermal expansion process. During the isothermal expansion of a classical ideal gas, free energy decreases, entropy increases and internal energy stays constant, see Fig. 7(a). In the presence of quantum confinement, similar behaviors are observed except internal energy does not remain constant but decreases, Fig. 7(b), due to the confinement energy coming from the quantum size effects\cite{sismanmuller,aydin1}. On the other hand, when a quantum confined gas effectively expands under SIST via a shape variable, free energy, entropy and internal energy can spontaneously and simultaneously decrease\cite{aydin7}, see Fig. 7(c). This counterintuitive behavior leads to some classically unexpected thermodynamic phenomena such as spontaneous transitions into lower entropy states, temperature-dependent work, isotropic heat and work exchanges, cooling by adiabatic compression or heating by adiabatic expansion, and work extraction occurring at the cold side (as opposed to hot side) of the cycles\cite{aydin7,aydinphd}. All of these extraordinary behaviors originate from the peculiar spectral changes due to SIST under quantum confinement at thermal equilibrium. Previously these behaviors have been examined from a more physical perspective such as the variation of quantum boundary layers due to temperature in confined domains\cite{aydin7,aydinphd}. From the thermodynamics perspective, in-phase behavior of internal energy, free energy and entropy causes many of these mentioned behaviors. Here, we go even further and understand the fundamental underlying reason for these unconventional behaviors by directly examining the characteristic changes in the energy spectrum as well as their effects on the thermal probabilities and properties.

\begin{figure*}[t]
\centering
\includegraphics[width=0.99\textwidth]{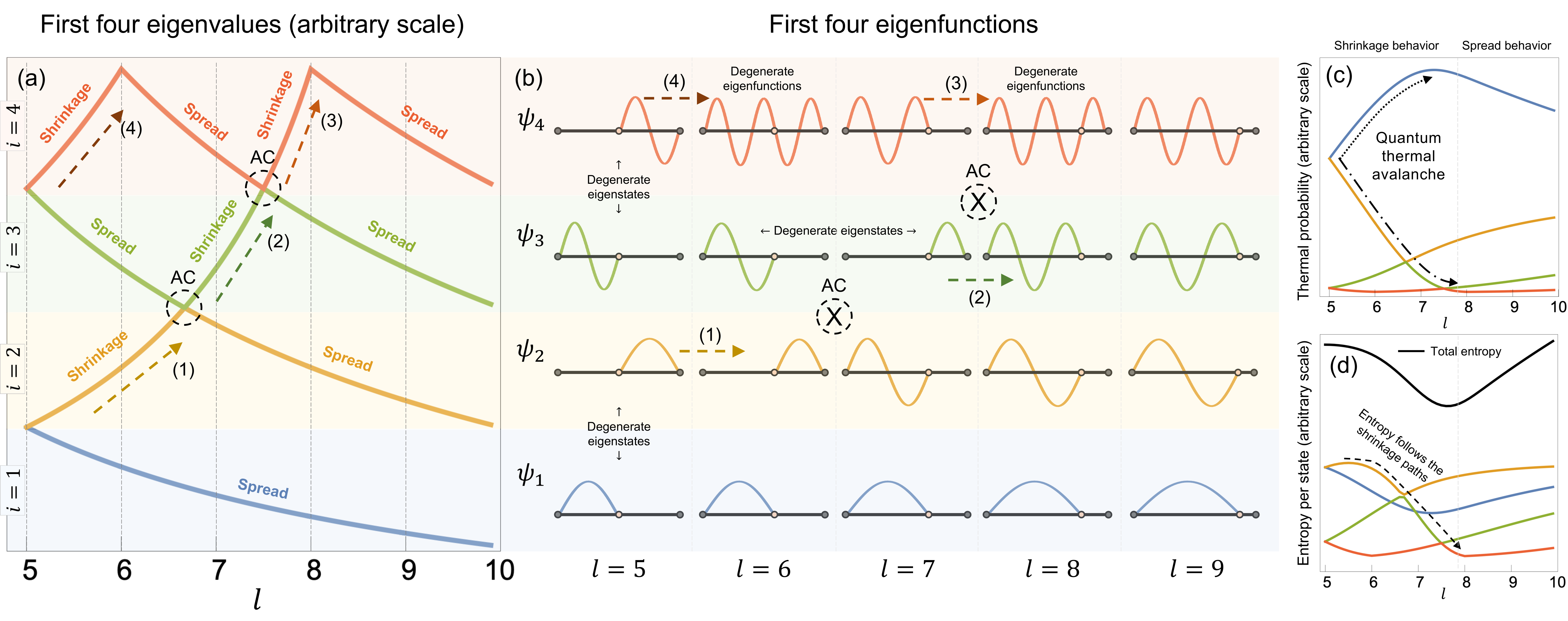}
\caption{Rush to the ground state: quantum thermal avalanche. (a) Behaviors of the first four eigenvalues with the inner boundary (partition) position $l$ in a 1D particle in a box. Partition is at the center for $l=5$ and at the right boundary of the box for $l=10$. (b) Behaviors of the first four eigenfunctions. Nonuniform level scaling is apparent from the mixed behaviors of spread and shrinkage of the eigenfunctions. Avoided crossings (AC) are also observed. (c) Thermal occupation probabilities of the first four states. Shrinkages of the excited state eigenfunctions cause swapping between ground state and excited state occupations leading to an abrupt increase in the ground state occupation probability, which we call quantum thermal avalanche. (d) Entropy contributions of first four states and the entropy shown with colored and black curves respectively.}
\label{fig:avalanche}
\end{figure*}

\section{Quantum thermal avalanche}

Even though the effective volume and quantum boundary layer concepts provide useful physical insights on how these type of unconventional changes occur in systems under SIST, by analyzing the eigenspectra, we can understand more about the nature of these changes as well as their physical origins. We examine the origins of the entropy anomalies in Fig. 8 in the eigenspectra and thermal probabilities. Columns represent the isothermal geometric operations in the considered systems, corresponding level spacings, behavior of the first few momentum eigenvalues, thermal probabilities, contributions of each states to the entropy and finally the variation of the entropy of the system with respect to the corresponding geometric control variable.

Increasing the length of a box uniformly decreases the eigenvalues, resulting a continuous decrease in the ground state occupation probability and successive increments in the excited state occupations, Fig. 8(a). As a result, entropy monotonically increases with the increasing length. Under SIST, however, changing the position of the partition causes a nonuniform change in the eigenvalues so that some eigenvalues rise while others descend and even undulate (going back and forth in a zig-zag movement), Fig. 8(b). As a result of these nonuniform changes, in certain parameter ranges the ground state occupation probability increases, while the thermal occupation probabilities of excited states exhibiting undulating behaviors. Due to the increased occupation of ground state as well as the accumulative behavior of undulating occupation probabilities of excited states, entropy decreases in the peculiar region despite the increase in effective volume, see the inset figures in the entropy subfigure. The winner of the competition between these two competing mechanisms essentially depends on the specific geometric configuration of the system. Similar behaviors are also observed for other SISTs having nonuniform level scaling such as the translational 2D (Fig. 8(c)) and rotational 2D (Fig. 8(d)) ones. Note that in all cases the geometric operations cause some sort of expansion as it is evident from the increase in the number of available states characterized by the increase in the partition functions shown in the insets of the entropy subfigures of Fig. 8.

Let's focus on the simplest one of the considered systems to understand the spontaneous entropy reduction. Consider the 1D particle in a box with moving partition (1D translational SIST), that is confined enough (e.g. the temperature is low or size is small enough) to be represented by the thermal probabilities of just the first four eigenvalues (for simplicity). In Fig. 9(a), we plot the variation of the first four eigenvalues $k_i$ with the shape variable $l$, position of the partition. In Fig. 9(b), we plot the variations of the corresponding eigenfunctions, $\psi_i$. We present the thermal probabilities $p_i$ in Fig. 9(c) and contributions to the entropy of each state $S_i$ in Fig. 9(d). Entropy is the sum of the contributions of all four states, which is shown with the black curve. At $l=5$, the partition starts at the center and the ground state is degenerate occupying both left and right compartments. With the movement of the partition to the right, ground state of the system ($i=1$) monotonically decreases with the increasing length of the left compartment where it sits, see $\psi_1$ in the blue row of Fig. 9(b). However, when we look at the behavior of the first excited state ($i=2$), we see an undulating (zig-zag) behavior as a manifestation of nonuniform level scaling. It initially increases, then at some point reverses its behavior and decreases just like the ground state does. As it is seen from the yellow row of Fig. 9(b), the first excited state eigenfunction squeezes into the smaller part of the domain between $l=5$ and $l=6$ constituting the ground state of the right compartment. However, at a point between $l=6$ and $l=7$, occupying the first excited state of the larger compartment becomes energetically more favorable than occupying the ground state of the smaller compartment and it switches into the left compartment where its wavelength can stay larger. In other words, $\psi_2$ first shrinks as the right compartment's ground state, then transfers itself to the left compartment's first excited state and spreads there with the right-moving partition. The similar shrinkage and spread behaviors of the eigenfunctions are seen in the higher eigenstates. Eigenstate swappings manifest themselves as avoided crossings in the eigenspectrum. Due to the idealistic nature of the chosen system, the crossings are infinitesimally avoided (still they don't cross each other). The avoided crossings are gaped in other less idealistic systems, see Fig. 8(c) and (d). 

Unlike in the usual size variation, in this case, expansion and contraction occur simultaneously in the different parts of the system. Shrinkage and spread behaviors of eigenfunctions directly determine the faith of the entropy under SISTs. Peculiarities in entropy arises from the shrinkage behaviors of the eigenfunctions, marked by the colored-dashed arrows. While the spreading of the eigenfunction along with the corresponding monotonic decrease of the eigenvalues is a common behavior due to the physics of the expansion, the shrinkage of some eigenfunctions in fact results to the increase of the corresponding eigenvalues. These changes in the spectra manifest themselves as opposite behaviors in the thermal occupation probabilities, Fig. 9(c). Because of the shrinkages thermal occupation probability of the ground state abruptly increase while that of the first excited state decreases and higher excited states show undulating behavior, see Fig. 9(c). This is a complete opposite behavior of an expanding system. We call this effect quantum thermal avalanche, due to the excessive occupation of the ground state occurring in quantum confined systems at the thermal equilibrium. Although the system is effectively expanding with moving partition, entropy decreases because of the sharp increase in the ground state occupation probability which makes the system less disordered. While there are more available states (as the partition function increases), most of them sit in the ground state not contributing to the entropy. This explains why the entropy decreases between $l=5$ and $l=8$. It basically follows the shrinkages of the eigenfunctions, see the dashed black arrows in Fig. 9(c) and (d) and compare them with the dashed colored arrows denoting the shrinkages. Notice how smoothly they represent the different behaviors with eigenstate swaps at the avoided crossings. On the other hand, between $l=8$ and $l=10$, all four eigenfunctions coherently spread out with the moving partition. This causes the recovery of the usual expansion behavior of the system in the thermal probabilities so that ground state occupation decreases and excited state occupations increase successively, Fig. 9(c). Hence, entropy increases between $l=8$ and $l=10$, Fig. 9(d).

Quantum thermal avalanche is a direct result of the nonuniform level scaling that is caused by the geometric level couplings due to SISTs. Despite the shrinkages and spreads occur simultaneously in the system (since one compartment gets smaller as the other gets larger), the shrinkage behaviors could dominate the overall behavior for the wide range of the control variable. In other words, the decrease in entropy due to the contraction of the smaller compartment dominates the increase in entropy due to the expansion of the larger compartment. As we mentioned earlier, this asymmetricity comes from the decay of occupation probabilities in the smaller part. Quantum thermal avalanche concept also explains the exponential increase of the partition function between $l=5$ and $l=8$, which is basically due to the enhanced ground state occupation.

Similar to the entropy, changes in free energy and internal energy can also be interpreted from the perspective of ground state reduction and quantum thermal avalanche. Internal energy is basically a sum over the energy levels times the corresponding thermal occupation probabilities. Free energy can also be written by factoring out the ground state and defining excited states over the level differences, which brings out the explicit ground state dependence as a separate term. Due to these explicit dependencies on the ground state, decrease of the first eigenvalue directly leads to the reduction of internal energy and free energy. The changes in their slopes in Fig. 6 directly come from the changing behavior of the ground state occupation probability shown in Fig. 9(c). The maximum of the ground state thermal occupation probability separates the region that is dominated by the shrinkages (quantum thermal avalanche) and the spreads (the usual expansion behavior). All of our explanations are valid for other types of SISTs as our interpretations can be applied to the subfigures shown in Fig. 8(c) and (d).

While all of these are inherently boundary effects, they are not seen in the classical systems because the energy quantization only becomes appreciable in quantum confined systems. Quantum thermal avalanche effect and all the other consequences of the SIST relies on the importance of the ground and low-lying states in determining the physical properties of the system. Otherwise, for larger (unconfined) systems with highly occupied states, these effects will be negligible, as we have shown in Fig. 3, since the spectral differences become indistinguishable.

\begin{figure}[t]
\centering
\includegraphics[width=0.48\textwidth]{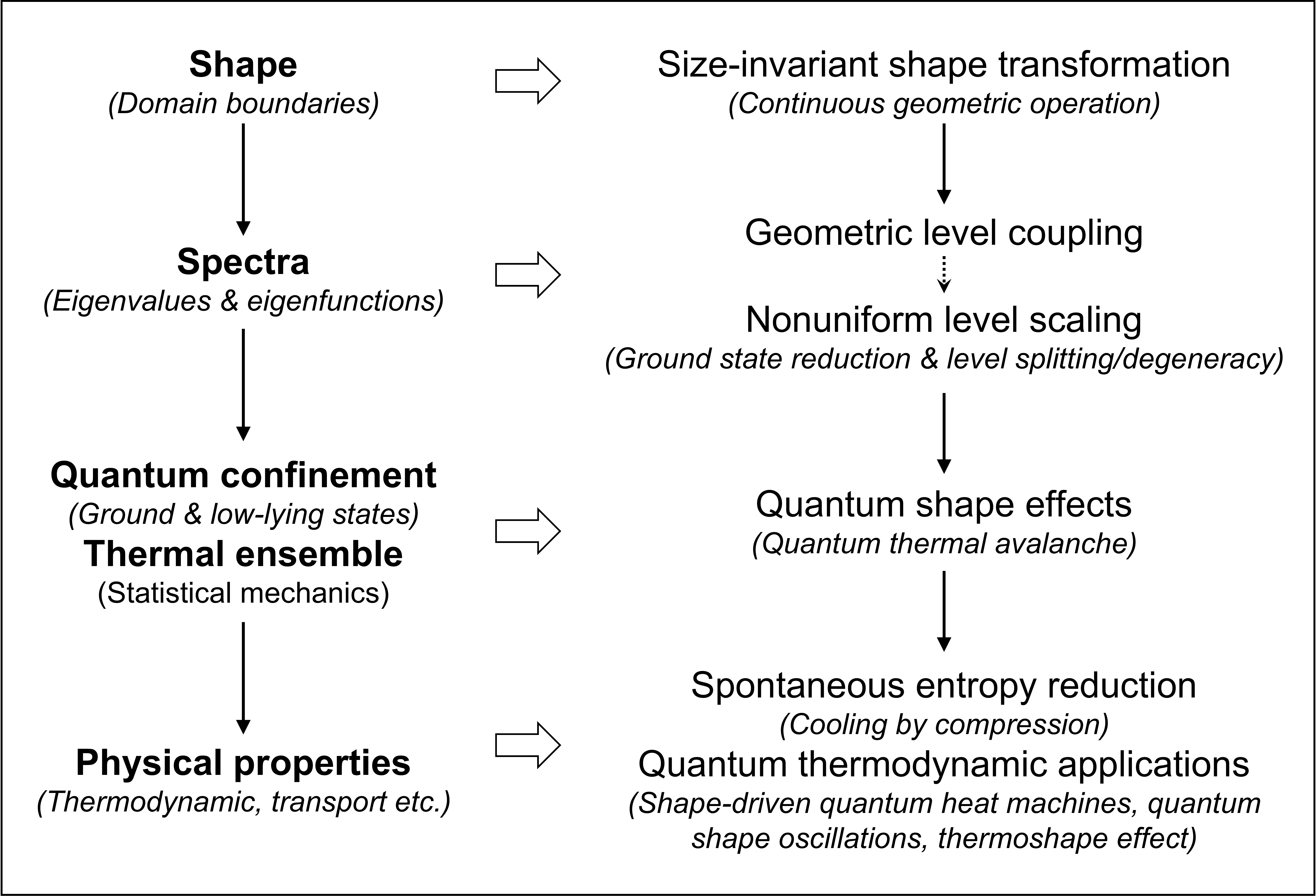}
\caption{Summary and physical consequences of the spectral signatures of size-invariant shape transformation.}
\label{fig:summary}
\end{figure}

\section{Conclusion}

In this paper we investigated the characteristic modifications in the Dirichlet spectra of various domains due to the boundary effects under a specific geometric transformation called SIST. There are two types of SIST: translational and rotational. Both cause similar counterintuitive thermodynamic behaviors. For convenience, we worked with 1D and 2D systems but our results hold in any dimension. In fact the three cases that we considered are quite general and comprehensive. By using the procedures of SIST, one can design arbitrary domains in any dimension exhibiting the similar behaviors. Our work emphasizes, perhaps once again, how subtle effects can appear when two boundaries come very close to each other at nanoscale systems.

We gave the general outline of our work in Fig. \ref{fig:summary}. Shape of a domain determines its spectra. A continuous geometric operation of SIST creates a particular geometric coupling between the levels which gives rise to a nonuniform scaling of levels. Quantum confinement leads to an increase in the importance of the ground state in determining the physical properties of the system. Hence, the ground state reduction and spectral gaps cause some behaviors that are not observed in classical systems, such as quantum thermal avalanche under the umbrella of quantum shape effects. As a result of these, peculiar behaviors are seen in thermodynamic properties such as the spontaneous transition into the lower entropy states. Quantum shape effects have many novel quantum thermodynamic applications, such as shape driven quantum heat machines\cite{aydin7,aydinphd}, quantum shape oscillations in thermodynamic properties of Fermionic systems\cite{aydin12} and thermoshape effect for energy harvesting with nanostructures\cite{aydin11}.

Designing nanostructures with almost arbitrary size and shapes is possible with the state-of-the-art nanotechnology\cite{1Dpotexp,Navon2021,scibox}. SISTs are therefore within the reach of existing experimental setups and utilizing the peculiar quantum thermodynamic features of SISTs could be considered as a viable direction for nanoscience and nano energy technologies. Our work could provide a theoretical basis on designing confinement geometries that could enhance the performance and efficiency of quantum thermal machines. Apart from nanostructures, our spectral findings could also be applied to the studies of quantum billiards, graphs as well as optical cavities. In fact, our work could basically have implications for all types of few-level systems. Beyond the spectral geometry, the ideas and findings we presented here could also provide new insights to the spectral theory in a broader sense.

\section*{Acknowledgments}

Author thanks to Asli Tuncer for useful discussions.

\appendix*
\section{Temperature-dependent effective volume and quantum boundary layer}

We used a concept called the effective volume to explain some of the thermodynamic behaviors. Here we give a brief information about the effective volume and quantum boundary layer concepts. More detailed examinations can be found in the Section 2.3 of Ref.\cite{aydinphd} A single quantum particle occupies the confined space in an inhomogeneous way even at thermal equilibrium due to its wave nature\cite{qbl}. Thermally weighted quantum probability density (quantum thermal density in short) of the confined particle is given by
\begin{equation}
n(T,\textbf{r})=\sum_i p_i(T)\left|\psi_i(\textbf{r})\right|^2,
\end{equation}
where $\textbf{r}$ is the position vector and $\psi_i$ is the eigenfunction of the $i$'th state. Eq. (A.1) contains both the thermal and quantum probabilities so that effectively capturing quantum-thermodynamic nature of the confined system. We used this equation to plot the quantum thermal densities in Figs. 7 and 8. 

As a result of the nonuniform density distribution even at thermal equilibrium, effectively empty regions form near the domain boundaries which is called the quantum boundary layer\cite{qbl,uqbl,nanocav}. The thickness of the quantum boundary layer for a confined ideal gas obeying Maxwell-Boltzmann distribution is universal (shape-independent) and exactly one-fourth of the thermal de Broglie wavelength, 
\begin{equation}
\delta(T)=\frac{\lambda_{th}(T)}{4},
\end{equation}
where $\lambda_{th}=\hbar\sqrt{2\pi}/\sqrt{mk_BT}$ is the thermal de Broglie wavelength of particles. 

Since the confined particles effectively occupy the rest of the domain that is not excluded by quantum boundary layers, one can introduce a new, effective volume by removing the quantum boundary layers from the actual volume. In the case of SISTs, quantum boundary layers can overlap which increases the available domain for particles to occupy\cite{aydin7}. Thereby, the effective volume is written as
\begin{equation}
\mathcal{V}_{\mli{eff}}(T,\gamma)=\mathcal{V}-\mathcal{V}_{\mli{qbl}}(T)+\mathcal{V}_{\mli{ovr}}(T,\gamma),
\end{equation}
where $\gamma$ is a generic shape variable. It has been shown that effective volume can accurately predict both quantum size and quantum shape effects\cite{aydin7,origin}. An interesting feature of the effective volume is its temperature dependence manifested in the thermal de Broglie wavelength. As a result, pressure or torque due to quantum confinement effects depends on temperature. This makes the work term in the first law of thermodynamics temperature dependent, giving rise to some classically impossible thermodynamic cycles as they are investigated in Refs.\cite{aydin7,aydinphd}.

\bibliography{main}
\bibliographystyle{unsrt}
\end{document}